# Terahertz radiation imaging of ferroelectric domain topology in room-temperature organic supramolecular ferroelectrics

M. Sotome[1], N. Kida[1], S. Horiuchi[2,3], and H. Okamoto[1]

[1]*Department of Advanced Materials Science, The University of Tokyo, 5-1-5 Kashiwa-no-ha, Chiba 277-8561, Japan*

[2]*National Institute of Advanced Industrial Science and Technology (AIST), Tsukuba 305-8562, Japan*

[3]*CREST, Japan Science and Technology Agency (JST), Tokyo 102-0075, Japan*

**Abstract**

Organic ferroelectrics with lightweight, flexible, low-cost, and environmentally benign characteristics are recently receiving great attention for new electric and optical devices. Since the propagation of ferroelectric domain walls and the subsequent reformation of ferroelectric domains are the basis for these devices, detection of the ferroelectric domain topology is crucial. Here, we demonstrate a new method to detect ferroelectric domains in inside and surface regions of organic ferroelectrics by mapping out two orthogonally polarized terahertz waves radiated from the crystal upon the irradiation of near-infrared femtosecond laser pulses. We used polarization dependence of the effective depths radiating the terahertz waves, which originate from the optical anisotropy in the terahertz frequency region. This allows us to distinguish ferroelectric domains in the inside and surface regions of the crystals. We applied this method to a room-temperature organic supramolecular ferroelectric crystal, 1:1 salt of 5,5'-dimethyl-2,2'-bipyridine and deuterated iodanilic acid. A single domain covering almost all the area of an as-grown crystal (~600 μm × 800 μm) is discerned in the inside region, while complicated multi-domain in size of ~200 μm is observed in the surface region. By applying external electric field along the 2*c-b* axis (ferroelectric polarization direction), the polarization switching proceeds with successive propagations of uncharged (neutral) and quasi-one-dimensional 180° domain walls (DWs) along the *b*-axis (⊥ 2*c-b* axis). This results in the formation of another uncharged and two-dimensional 180° DW parallel to the (100) plane, which covers all the area of the crystal. We discuss the usefulness of the present terahertz radiation imaging technique and ferroelectric DW dynamics in terms of anisotropic stacking of hydrogen-bonded chains.

# I. Introduction

In ferroelectrics, polarization switching by external electric fields proceeds by propagation of ferroelectric domain walls (DWs), which separate ferroelectric domains with different polarizations [1, 2]. Such a polarization switching process and the subsequent reformation of ferroelectric domains are the basis for electric and optical devices using ferroelectrics such as non-volatile memories, actuators, and nonlinear optical elements [1]. Thus, understanding of ferroelectric DW dynamics is needed not only to realize these devices but also to improve performance of existing devices. Ferroelectric domain topology as well as DW dynamics are determined by the symmetry of the crystal [2, 3], but in reality they are governed by various factors such as depolarization field at the surface [4, 5] and growth condition of the crystal [2], resulting in unpredictable complex patterns of ferroelectric domains and DWs [2]. DWs are roughly classified into two types in terms of relative angle between the electric polarization $P$ and DW [6], i.e., uncharged (neutral) DWs parallel to $P$ and charged DWs perpendicular to $P$. Since uncharged DWs are electrically stable (div$P$ = 0), they are observed in various ferroelectrics [2]. On the other hand, charged DWs such as head-to-head (tail-to-tail) DWs have net bound charge (div$P \neq 0$), which should be compensated by charged objects such as free carriers, defects, and impurities. Charged DWs frequently appear in as-grown crystals [7-10] and as-grown thin films [11-13]. Therefore, real-space imaging of ferroelectric domain topology is not only indispensable to understand the nature of ferroelectricity but also crucial for the applications of ferroelectric materials.

Recently, new series of room-temperature hydrogen-bonded ferroelectrics with large ferroelectric polarization have been developed in a family of low-dimensional organic molecular crystals [14-16]. Molecules in these crystals are connected by hydrogen bonds; cooperative proton transfer along with the asymmetric π-electron conjugation change induces ferroelectric polarization. Compared to inorganic ferroelectrics, organic ferroelectrics have advantages in lightweight, flexible, low-cost, and environmentally benign characteristics. Thus, they are now receiving great research attention in terms of new room-temperature electronic and photonic applications. In the context of these applications, the detection of ferroelectric domain topology is indispensable. Recent works on a croconic acid (4.5-dihydroxy-4-cyclopentene-1,2,3-trion) [17] and [H-6,6'-dmbp][Hca] (6,6'-dimethyl-2,2'-bipyridinium chloranilate) [18], both of which are room-temperature hydrogen-bonded ferroelectrics, have revealed the presence of charged head-to-head (tail-to-tail) DWs in the virgin sate of as-grown crystals. The charged DWs diminish in electrically poled crystals [17] or thermally annealed crystals [18]. In the study of a croconic acid [17], ferroelectric domain topology was detected by mapping out the amplitude of terahertz waves radiated from the crystal upon the irradiation of femtosecond laser pulses. Since the thickness or depth inside the crystal contributing to the terahertz wave radiated outside the crystal was long (~600 μm), the terahertz radiation image obtained in a 100-μm-thick sample reflects bulk ferroelectric domains. In the study of [H-6,6'-dmbp][Hca], on the other hand, surface-sensitive piezoresponce force microscopy (PFM) was used to detect the ferroelectric domain topology [18]. In general, domain topology is different in the surface and inside regions of the crystal. PFM detects ferroelectric domains in the surface, while the terahertz radiation imaging method applied on croconic acid [17] detects bulk ferroelectric domains. These methods provide complementary information on the ferroelectric domain topology. Another important difference in the two methods is the size of measurable areas. In PFM study of [H-6,6'-dmbp][Hca], typical measured area was ~5 μm × 5 μm [18]. On the other hand, in terahertz radiation imaging study of croconic acid, the measured area was ~400 μm × 400 μm, which covers all a crystal plane of as-grown single crystal [17]. In order to further reveal the characteristics of the ferroelectric domain topology in organic ferroelectrics, it is necessary to develop a method, which can distinguish orientations of ferroelectric domains in the inside and surface regions covering the whole crystal volume. However, this activity has not been carried out for organic ferroelectrics so far.

Here, we propose a new method to visualize ferroelectric domains and DWs in the inside and surface regions of organic ferroelectrics and applied this method to an organic molecular crystal, 1:1 salt of 5,5'-dimethyl-2,2'-bipyridine (55DMBP) and deuterium iodanilic acid ($D_2ia$) ([D-55DMBP][Dia]), which was recently found to be a room-temperature ferroelectric [19]. We show that terahertz wave is emitted by the irradiation of a femtosecond laser pulse and effective depth radiating the terahertz waves exhibits large polarization dependence. By mapping out the electric field amplitude of two orthogonally polarized terahertz waves, we successfully visualize distinct ferroelectric domains in the inside and surface regions of the crystal. In the inside region of an as-grown crystal, an almost single domain is discerned, while multi-domain is observed in the surface region. The polarization switching by an external electric field can be explained by successive propagations of uncharged and quasi-one-dimensional (1D) 180° DWs along the direction perpendicular to the ferroelectric polarization within two-dimensional (2D) molecular chain layers. This results in the formation of interlayer 2D 180° DW almost parallel to the (100) plane. On the basis of systematic optical measurements in the terahertz and visible frequency regions, we discuss DW dynamics and principles of the present terahertz radiation imaging technique.

[D-55DMBP][Dia] investigated here is deuterated analog of 1:1 salt of 5,5'-dimethyl-2,2'-bipyridine (55DMBP) and iodanilic acid ($H_2$ia) ([H-55DMBP][Hia]) with Curie temperature $T_c$ = 268 K [19]. Crystal structure of [H-55DMBP][Hia] is triclinic with the space group of $P1$ and $P\bar{1}$ in the ferroelectric and paraelectric phase, respectively. 55DMBP and $H_2$ia molecules arrange alternately to form one-dimensional hydrogen-bonded chains parallel to the $2c$-$b$ axis, as shown in Fig. 1. Protons in the intermolecular hydrogen bonds locate off-center positions, which is accompanied by asymmetric π–electron distribution, and thus induces electric polarization along the hydrogen-bonded chain axis, i.e., $2c$-$b$ axis [19]. The positions of the protons can be switched by an external electric field (Fig. 1), resulting in the polarization reversal. A notable characteristic of [H-55DMBP][Hia] is the deuterium effect on the ferroelectric properties; deuterated single crystal, [D-55DMBP][Dia], shows the spontaneous polarization $P_s$ of ~2 μC/cm$^2$ even at room temperature ($T_c$ = 335 K) [19]. In this work, we focused on [D-55DMBP][Dia].

## II. Experimental Procedure

(100)-oriented single crystals of [D-55DMBP][Dia] were obtained by recrystallization from a solution. Details of the sample preparation were reported elsewhere [19].

Polarized transmittance spectra in the terahertz frequency region were measured by the standard terahertz time-domain spectroscopy [20]. We used a 0.5-mm-thick (110)-oriented ZnTe single crystal as a terahertz wave emitter and photoconducting switch on a low-temperature-grown GaAs substrate as a detector. The 0.15-mm-thick (100)-oriented single crystal of [D-55DMBP][Dia] was attached on the holder with a pinhole; the diameter of the pinhole was 1 mm.

We measured polarized transmittance and reflectance spectra in the energy range of 0.7-5.0 eV by a grating spectrometer.

In the terahertz radiation experiments [17, 20], femtosecond laser pulses delivered from a mode-locked Ti:sapphire laser (the wavelength of 800 nm, the repetition rate of 80 MHz, and the pulse width of 100 fs) was focused on a sample in normal incidence. Unless otherwise stated, laser power was set to be 5 mW. The spot diameter of the laser pulse was estimated to be ~25 μm, which corresponds to 25 μJ/cm$^2$ per pulse. We employed photoconductive sampling technique to measure the electric-field waveform of the radiated terahertz waves [20]. Two wire grid polarizers (WG1, WG2) were used to confine the detected polarization of the radiated terahertz waves. A pair of electrodes with conducting carbon paste was attached on the side (102) surfaces of a crystal so as to apply the electric field along the 2*c*-*b* axis. Further detailed experimental procedures are described in Sects. III. B and III. C.

All the experiments were performed at room temperature.

## III. Results and Discussion
## A. Polarized optical spectra in terahertz and visible frequency regions

First, we detail the polarized reflectance *R* and transmittance *T* spectra in the energy range of 0.7-5.0 eV. *R* spectra obtained with the electric field of light $E$ parallel to *b* axis ($E \parallel b$) and 2*c*-*b* axis ($E \parallel 2c - b$) are shown in the right panel of Fig. 2(a) by orange and green solid lines, respectively. No remarkable anisotropy is observed. In both polarizations, strong peak structures are discerned at ~3.8 eV, assigned to intramolecular electronic excitations, i.e., π-π* transitions of the molecules. We also show in Fig. 2(a) polarized $T$ spectra by orange and green dotted lines. The steep decrease of $T$ above 1.7 eV is due to the presence of π-π* transitions. We performed the Kramers-Kronig transformation of $R$ spectra to extract the refractive index $n$ spectra in $E \parallel b$ and $E \parallel 2c - b$ configurations, which are shown in Fig. 2(b). The $n$ spectra in the energy range of 1.0-2.0 eV were reproduced by the Sellmeier relationship, which is represented by $n = \sqrt{1 + \frac{S_0 \lambda_0^2}{1 - \left(\frac{\lambda_0}{\lambda}\right)^2}}$, where $\lambda$ is the wavelength. This yields $S_0 = 5.17 \times 10^{-5}$ nm$^{-2}$ and $\lambda_0 = 193$ nm for $E \parallel b$, while $S_0 = 5.53 \times 10^{-5}$ nm$^{-2}$ and $\lambda_0 = 190$ nm for $E \parallel 2c - b$. With the use of the obtained fitting parameters, we estimated the group refractive index $n_g$ spectra, which is given by $n_g = |n - \lambda \frac{dn}{d\lambda}|$, as shown in Fig. 2(b). There is little anisotropy of $n_g$ at 1.55 eV (the photon energy of the femtosecond laser pulses in this study), which was evaluated to be ~1.8 for both configurations. Figure 2(c) shows the absorption coefficient $\alpha$ spectra; solid lines were obtained by the Kramers-Kronig transformation of the $R$ spectra, while dotted lines were derived from the relation $\alpha = -\frac{1}{d} \ln \left( \frac{T}{(1-R)^2} \right)$ with $d$ the thickness of the sample ($d$ = 0.15 mm). $\alpha$ at 1.55 eV was estimated to be ~10 cm$^{-1}$, which corresponds to a penetration depth of ~1 mm for the femtosecond laser pulses in terahertz radiation experiments (Sect. III. B).

In the left panel of Fig. 2(a), we show the polarized transmittance $T$ spectra in the terahertz frequency region in $E \parallel b$ and $E \parallel 2c - b$ configurations. The measured frequency ranges in $E \parallel b$ and $E \parallel 2c - b$ configurations were 0.7-3.0 THz and 0.7-2.2 THz, respectively. In contrast to $T$ spectra in the visible frequency region, there is a large anisotropy; $T$ at 1.0 THz in the $E \parallel 2c - b$ configuration is ~100 times smaller than that in the $E \parallel b$ configuration. In the $E \parallel b$ configuration, we identify peak structures at ~1.6 THz and ~2.4 THz, which would be assigned to intramolecular or intermolecular vibrations. On the other hand, broad absorption in the measured frequency region (0.7-2.2 THz) is discerned in the $E \parallel 2c - b$ configuration. From the measured $T$ spectra, we extracted $n$ and $\alpha$ spectra, which are shown in Figs. 2(b) and 2(c), respectively. Details of our derivation procedure are described in Ref. [22]. $n$ in the $E \parallel 2c - b$ configuration increases with lowering frequency, while $n$ in the $E \parallel b$ configuration shows the dispersion of multiple Lorentz oscillators. Accordingly, relatively large $\alpha$ in the $E \parallel 2c - b$ configuration is discerned over the whole measured frequency range. This broad absorption was previously reported in the related analog, [H-55DMBP][Hia], by far-infrared reflectivity measurements [21]; its origin was discussed in terms of the enhancement of the proton fluctuation.

**B. Terahertz radiation**

In terahertz radiation experiments, we used a 240-μm-thick (100)-oriented single crystal. Since the penetration depth (~1 mm) of the incident femtosecond laser pulses exceeds the thickness of the sample, attenuation of the laser pulse inside the sample is negligible. External electric fields $E_{\text{ex}} = \pm 5\,\text{kV/cm}$, which exceed the coercive field (~2 kV/cm) [15], were applied along $2c - b$ axis. The electric fields of the femtosecond laser pulses $E^\omega$ and the detected terahertz waves $E_{\text{THz}}$ were set parallel to $2c - b$ axis, i.e., $E^\omega \| 2c - b$ and $E_{\text{THz}} \| 2c - b$. Figure 3(a) shows the terahertz electric-field waveforms $E_{\text{THz}}(t)$ radiated from the crystal. A nearly single-cycle pulse around 0 ps is discerned in a time window from -0.6 ps to 0.4 ps and subsequent temporal oscillation components emerge up to ~8 ps. We observed reversal of the terahertz waveform when the sign of $E_{\text{ex}}$ was changed. As detailed in the Sec. III. D, this is caused by a polarization switching [the inset of Fig. 3(a)]. To see the spectra of the terahertz wave, we performed fast Fourier transformation (FFT) on the measured waveforms, the results of which are shown in Fig. 3(b). They contain frequency components up to 3 THz and have two dip structures at around 1.2 THz and 1.7 THz. The origins of these dip structures are unclear at present, since a broad absorption in the $E||2c-b$ configuration obscures other spectral features such as optical phonons [Fig. 2(c)].

We also observed emission of terahertz waves in the $E^\omega||b$ and $E_{\text{THz}}||b$ configuration; typical terahertz electric-field waveforms with $E_{\text{ex}} = \pm 5$ kV/cm and their FFT spectra are shown in Figs. 3(c) and 3(d), respectively. The electric field at 0 ps in this configuration [Fig. 3(c)] is slightly (1%) larger than that in the $E^\omega||2c-b$ and $E_{\text{THz}}||2c-b$ configuration [Fig. 3(a)]. The spectral width of the terahertz waveform is narrower than that obtained in the $E^\omega||2c-b$ and $E_{\text{THz}}||2c-b$ configuration and the temporally oscillation component is more pronounced, which appears as the peaks in the frequency range of 1-2 THz [Fig. 3(d)]. It was also found that the phase of the radiated terahertz waves is sensitive to the direction of the ferroelectric polarization, as illustrated in the inset of Fig. 3(c).

So far, various optical processes have been proposed as the terahertz radiation mechanism in non-centrosymmetric media [20]. In room-temperature hydrogen-bonded organic ferroelectrics such as croconic acid [17] and 2-phenylmalondialdehyde [23], the terahertz radiation mechanism is optical rectification of the incident laser pulse by second-order nonlinear optical process [24]. Terahertz radiation mechanisms related to photocarrier generation can be ignored for [D-55DMBP][Dia], since it is transparent to the near-infrared femtosecond laser pulse. Thus, possible terahertz radiation mechanisms are optical rectification [24] and/or coherent phonon generation by impulsive stimulated Raman scattering [22,25-27]; both processes can be described by second-order nonlinear optical coefficient tensor $\chi^{(2)}_{ijk}$. A laser-induced nonlinear polarization $P^{\mathrm{NL}}$ is given by

$$P_i^{\mathrm{NL}} = \epsilon_0 \chi^{(2)}_{ijk} E_j^\omega E_k^{\omega *}, \quad (1)$$

where $\epsilon_0$ is the vacuum permittivity. All the $\chi^{(2)}_{ijk}$ components are allowed in the space group of $P1$ [28].

To reveal the role of the second-order nonlinear optical effects on the emission of terahertz waves, we measured the laser polarization and power dependences. Figure 4(a) shows schematic illustration of the experimental setup for the measurement of the laser polarization dependence. We used a 240-μm-thick (100)-oriented single crystal in this experiment. Single-domain state was obtained by application of $E_{\mathrm{ex}} = 5$ kV/cm along 2*c*-*b* axis. The laser power was set to be 10 mW. We rotated polarization of the incident femtosecond laser pulses by the angle $\theta$ with a half-wave plate; $\theta$ was defined as the angle of the laser polarization relative to the *X*-axis. We measured terahertz waveforms $E_{+45°}(t)$ and $E_{-45°}(t)$ by setting WG1 angle +45° and -45°, respectively [29]. Terahertz waveforms of the horizontal ($X \parallel 2c - b$) component $E_X(t)$ and vertical ($Y \parallel b$) component $E_Y(t)$ were obtained from $E_{+45°}(t)$ and $E_{-45°}(t)$ by the relation

$$E_X(t) = \frac{1}{2}(E_{+45°}(t) + E_{-45°}(t)), \quad (2)$$

$$E_Y(t) = \frac{1}{2}(E_{+45°}(t) - E_{-45°}(t)). \quad (3)$$

Amplitude spectra of the waveforms $E_X(\omega)$ and $E_Y(\omega)$ were obtained by FFT of $E_X(t)$ and $E_Y(t)$, respectively. When the light-induced nonlinear polarization given by Eq. (1) is dominant in the terahertz radiation process, power spectra of the terahertz waves $|E_i(\omega)|^2$ ($i = X, Y$) in terms of $\theta$ is given by

$$|E_i(\omega)|^2 \propto \left[\chi^{(2)}_{iXX}(\omega)\cos^2\theta + \left(\chi^{(2)}_{iYX}(\omega) + \chi^{(2)}_{iXY}(\omega)\right)\cos\theta\sin\theta + \chi^{(2)}_{iYY}(\omega)\sin^2\theta\right]^2 I_0^2 L_{\text{gen}}^2(\omega) T'(\omega), \quad (4)$$

where $I_0$ is the incident laser power, $L_{\text{gen}}$ is the effective generation length [22,30], $T'(\omega) = \frac{4n^2}{(n+1)^2 + \kappa^2}$ is the transmittance of the terahertz waves at the sample-air interface with $\kappa$ the extinction coefficient. $L_{\text{gen}}$ represents the effective length of the region contributing to the observed terahertz waves; $L_{\text{gen}}$ is given by

$$L_{\text{gen}}(\omega) = \left(\frac{1 + \exp(-\alpha(\omega)d) - 2\exp\left(-\frac{\alpha(\omega)}{2}d\right)\cos\left(\frac{\omega}{c}|n(\omega) - n_g|d\right)}{\left(\frac{\alpha(\omega)}{2}\right)^2 + \left(\frac{\omega}{c}\right)^2 (n(\omega) - n_g)^2}\right)^{\frac{1}{2}}, \quad (5)$$

where $c$ is speed of light in vacuum.

Here, $|E_i(\omega)|^2$ was integrated in the frequency range of 0.8-1.2 THz, where the terahertz radiation is relatively strong for both polarizations [Figs. 3(b) and 3(d)]. $T'$ in $E_{\text{THz}}||2c - b$ and $E_{\text{THz}}||b$ configurations were 0.68 and 0.89, respectively. $L_{\text{gen}}$ of the 240-μm-thick crystal were 28 μm ($E_{\text{THz}}||2c - b$) and 205 μm ($E_{\text{THz}}||b$). Figure 4(b) shows the measured $\theta$ dependence of $|E_i(\omega)|^2$. Equation (4) reproduces the

measured $\theta$ dependence of $|E_i(\omega)|^2$, as indicated by solid lines in Fig. 4(b). The resultant ratios of $\chi^{(2)}_{ijk}$ components to $\chi^{(2)}_{YYY}$ are as follows; $\chi^{(2)}_{XXX}/\chi^{(2)}_{YYY} = -8.3$, $\chi^{(2)}_{XYY}/\chi^{(2)}_{YYY} = 5.3$, $\chi^{(2)}_{YXX}/\chi^{(2)}_{YYY} = -0.7$, $(\chi^{(2)}_{YYX} + \chi^{(2)}_{YXY})/\chi^{(2)}_{YYY} = -0.7$, and $(\chi^{(2)}_{XYX} + \chi^{(2)}_{XXY})/\chi^{(2)}_{YYY} = 5.1$. We show in Fig. 4(c) the laser power dependence of $|E_X(\omega)|^2$ and $|E_Y(\omega)|^2$ in the $E^\omega || 2c - b$ configuration, which are also well reproduced by Eq. (4) [solid lines in Fig. 4(c)]. These results indicate that terahertz radiation in [D-55DMBP][Dia] is induced by second-order nonlinear optical process. Since sharp dip structures, which may be related to the phonon modes, are observed in the power spectra of the terahertz radiation [Figs. 3(b) and 3(d)], the comprehensive Raman scattering experiments are needed to further discuss the terahertz radiation mechanism. Here, we focus on the ferroelectric domain imaging using emission of terahertz wave.

## C. Ferroelectric domain imaging using emission of terahertz waves

In order to visualize ferroelectric domains and DWs [3], various methods have been developed so far. Among them, optical means were widely used to detect ferroelectric domains over a wide area of the sample [2, 31]. Using emission of terahertz waves, spatial distribution of ferroelectric domains and DWs have been visualized in various ferroelectrics such as $BiFeO_3$ thin film [32], $Co_3B_7O_{13}I$ [33], and croconic acid [17]. It was found that the phase of the terahertz wave radiated from ferroelectric materials upon the irradiation of a femtosecond laser pulse depends on the direction of ferroelectric polarization. Thus, ferroelectric domains with different polarizations can be easily distinguished by mapping out the amplitude of the radiated terahertz waves over the sample. This method was recently applied to visualize ferroelectric domains with different polarizations in croconic acid [17] and $Co_3B_7O_{13}I$ [33]. In croconic acid, tail-to-tail and 180° domains were successfully detected in as-grown and electrically poled crystals, respectively [17]. On the other hand, 90° domain was discerned in $Co_3B_7O_{13}I$ [33]. In previous studies on croconic acid [17] and $Co_3B_7O_{13}I$ [33], we mapped out the electric field of the terahertz wave polarized only along the ferroelectric polarization direction. Since the detected signal of ferroelectric domains along the depth direction is averaged, we were not able to detect three-dimensional ferroelectric domain and DWs in the crystals.

On the other hand, a notable feature of the terahertz radiation in [D-55DMBP][Dia] is that two orthogonally polarized terahertz waves can be generated (Fig. 3). In addition, this compound is optically anisotropic and absorption coefficient along $2c-b$ is much larger than along $b$ in the terahertz frequency region (Fig. 2). This makes us expect that different parts in the depth direction contribute to the radiated terahertz waves with those two polarization directions $E_{\text{THz}}||2c-b$ and $E_{\text{THz}}||b$. Using this, we will be able to evaluate the depth profiles of the contributions to the radiated terahertz wave and therefore obtain the information of the direction of the ferroelectric polarization. The amplitude spectrum $E_{\text{THz}}(\omega, d)$ of the terahertz wave at the far point outside the crystal is given by

$$E_{\text{THz}}(\omega, d) = H(\omega) \frac{\mu_0 \chi^{(2)}(\omega) \omega^2 I(\omega)}{n_o \left\{ \frac{c}{\omega} \left[ \frac{\alpha(\omega)}{2} + \alpha_o \right] + i[n(\omega) + n_g] \right\}}$$

$$\times \frac{\exp\left[-\frac{i\omega n(\omega)}{c} d\right] \exp\left[-\frac{\alpha(\omega)}{2} d\right] - \exp\left[-\frac{i\omega n_g}{c} d\right] \exp[-\alpha_o d]}{\frac{\alpha(\omega)}{2} - \alpha_o + \frac{i\omega}{c}[n(\omega) - n_g]}, \quad (5)$$

where $\mu_0$ is magnetic permeability in vacuum, $\omega_0$ is the central angular frequency of the femtosecond laser pulse (800 nm~375 THz), $\alpha_0$ is the absorption coefficient at $\omega_0$ (~9 cm$^{-1}$ for $E_{\text{THz}}||2c-b$ and ~8 cm$^{-1}$ for $E_{\text{THz}}||b$), $n_o$ is the refractive index at $\omega_0$ (~1.78 for $E_{\text{THz}}||2c-b$ and ~1.75 for $E_{\text{THz}}||b$), and $I(\omega)$ is the Fourier-transformed intensity spectrum of the femtosecond laser pulse [22,30]. Here, $H(\omega)$ is detection response function of the LT-GaAs detector used in our experiments, as detailed in Ref. [22]. We derived $\chi^{(2)}$ spectra in $E_{\text{THz}}||2c-b$ and $E_{\text{THz}}||b$ configurations from Eq. (5) using FFT amplitude spectra of the terahertz waves shown in Figs. 3(a) and 3(c), respectively. Frequency dependent contribution $G(\omega,z)$ of the terahertz waves generated at depth $z$ to the observed $E_{\text{THz}}(\omega,d)$ is given by

$$G(\omega,z) = \chi^{(2)}(\omega)E_{\text{Inst}}(\omega)\exp\left[-\frac{i\omega n_g}{c}z\right]\exp[-\alpha_o z]$$
$$\times \exp\left[-\frac{i\omega n(\omega)}{c}(d-z)\right]\exp\left[-\frac{\alpha(\omega)}{2}(d-z)\right], \quad (6)$$

where $E_{\text{Inst}}(\omega) = \omega^2 H(\omega)I(\omega)$ is the instrumental function in our experimental setup, as detailed in Ref. [22]. Here, we focus on the electric field of the terahertz waveform at 0 ps. Thus, spectrally integrated $G(\omega,z)$ gives the depth profile of the contribution to the detected terahertz wave $G(z)$, which is given by

$$G(z) = \text{Re}\left\{\int_{\omega_1}^{\omega_2} d\omega G(\omega,z) \Big/ \int_{\omega_1}^{\omega_2} d\omega G(\omega,d)\right\}. \quad (7)$$

Since the back surface region mainly contributes to $E_{\text{THz}}(0)$, we divided $\int_{\omega_1}^{\omega_2} d\omega G(\omega,z)$ by $\int_{\omega_1}^{\omega_2} d\omega G(\omega,d)$. We obtained polarized $T$ spectra in the frequency range of 0.8-2.2 THz and 0.8-3.0 THz in $E_{\text{THz}}||2c-b$ and $E_{\text{THz}}||b$ configurations, respectively [Fig. 2(a)]. Thus, we set $\omega_1/2\pi$ and $\omega_2/2\pi$ the lowest and highest values of the measured frequency range, respectively. The calculated $G(z)$ profiles in $E_{\text{THz}}||2c-b$ and $E_{\text{THz}}||b$ configurations with $d = 150$ μm are shown in Figs. 5(a) and 5(b), respectively. We defined the effective depth $L_{\text{eff}}$ as the length from the back surface where $G(z)$ becomes $1/e$ of $G(d)$. In the $E_{\text{THz}}||2c-b$ configuration, $L_{\text{eff}}$ was estimated to be ~29 μm. This indicates that the back side region of the crystal with a depth of ~29 μm mainly contributes to the detected terahertz waves, as schematically shown in the inset of Fig. 5(a). On the other hand, $L_{\text{eff}}$ in the $E_{\text{THz}} \| b$ configuration reaches ~89 μm. Thus, terahertz waves generated from the inside region of the crystal are detected [inset of Fig. 5(b)].

In the terahertz imaging experiments, we measured the electric field of terahertz wave at 0 ps [$E_{\text{THz}}(0)$] at various positions of an as-grown crystal by a raster scan. The spot diameter of the incident femtosecond laser pulse, which corresponds to the spatial resolution of the terahertz radiation images, was ~25 μm. This experiment is conducted on a 150-μm-thick (100)-oriented as-grown crystal. First, we adopt the configuration shown in the lower part of Fig. 5(d), in which the crystal is irradiated with a femtosecond laser pulse from the front side, and $E^\omega$ and $E_{\text{THz}}$ were parallel to $2c - b$ axis. In this configuration, ferroelectric domains in the back surface region (~29 μm) can be visualized, since $L_{\text{eff}}$ is ~29 μm [Fig. 5(a)]. The upper panel of Fig. 5(d) shows the terahertz radiation image of the crystal. The optical microscope image of the same area is shown in Fig. 5(c). Although the crystal has a wide (100)-oriented flat surface, edge areas are not normal to the (100)-surface, as seen in Fig. 5(c); flat region of the crystal and thus actual valid area of the terahertz radiation image are surrounded by the dotted lines. The color scale shown in the left-hand side of Fig. 5(d) represents the magnitude and sign of $E_{\text{THz}}(0)$. The red and blue regions indicate that the sign of $E_{\text{THz}}(0)$ is different. Thus, these regions correspond to ferroelectric domains with opposite polarization directions, i.e., $P_s \| \pm (2c - b)$. In the white region, $E_{\text{THz}}(0)$ is nearly zero, which corresponds to the ferroelectric DWs. Since a width of each white region is almost equal to the spatial resolution (~25 μm), the actual DW width is narrower than ~25 μm. From this image, we can see that ferroelectric domains with size larger than ~200 μm exist in the back surface region with ~29 μm. These domains are separated by two types of DWs; uncharged 180° DWs with a plane parallel to $\pm(2c - b)$ axis and charged head-to-head (or tail-to-tail) DWs with a plane perpendicular to $\pm(2c - b)$ axis. In addition, we discern the presence of inclined DWs.

To reveal the ferroelectric nature inside the crystal, we measured terahertz radiation image in the $E_{\text{THz}}||b$ configuration, which is shown in Fig. 5(e). Since the terahertz radiation efficiencies in $E_{\text{THz}}||2c - b$ [Fig. 5(d)] and $E_{\text{THz}}||b$ [Fig. 5(e)] configurations are different, it is necessary to correct the value of $E_{\text{THz}}(0)$ for $E_{\text{THz}}||b$. In the case of the 240-μm-thick crystal, $E_{\text{THz}}(0)$ for $E_{\text{THz}}||b$ is found to be slightly (1%) larger than that for $E_{\text{THz}}||2c - b$ [Figs. 3(a) and (c)]. The $d$ dependence of $E_{\text{THz}}(0)$ is given by $\int_0^d dzG(z)$. Compared to the 240-μm-thick crystal, $\int_0^d dzG(z)$ of the 150-μm-thick crystal is 105% and 84% in $E_{\text{THz}}||2c - b$ and $E_{\text{THz}}||b$ configurations, respectively. Thus, $E_{\text{THz}}(0)$ in the $E_{\text{THz}}||b$ configuration was multiplied by $1.24 = 1.05/(0.84 \times 1.01)$ as the correction of the anisotropy in the terahertz radiation efficiency. In the $E_{\text{THz}}||b$ configuration, ferroelectric domains inside the crystal can be visualized, since $L_{\text{eff}}$ is ~89 μm [Fig. 5(b)]. Compared to the ferroelectric domain image shown in Fig. 5(d), different patterns of ferroelectric domains are observed in Fig. 5(e). The crystal consists of ferroelectric domains with opposite polarizations, a large blue-colored domain and a small red-colored domain, which is separated by a head-to-head (or tail-to-tail) DW. Note that the lower region of the crystal is occupied by the blue-colored domain, although the same area in the back surface region consists of the red-colored domain [Fig. 5(d)]. This indicates the presence of a quasi-2D 180° DW along the (100) plane in the lower region of the crystal.

We also performed a similar experiment by reversing the crystal along *b*-axis. Namely, the crystal is irradiated with a femtosecond laser pulse from the back side, as illustrated in the lower part of Fig. 5(f). The obtained terahertz radiation image is shown in the upper panel of Fig. 5(f). The ferroelectric domain image is similar to that shown in Fig. 5(e). This ensures that the terahertz radiation images obtained in the $E_{\text{THz}}||b$ configurations reflect the bulk ferroelectric domains.

Figure 5(g) shows the terahertz radiation image of the front surface region of the crystal measured in the $E_{\text{THz}}||2c-b$ configuration and with a femtosecond laser pulse incident from the back side. As can be seen, observed multi-domain structure is different from the images in the back surface and inside region [Figs. 5(d), (e) and (f)]. Ferroelectric domain indicated by the red region can be seen as white regions in bulk ferroelectric domain images shown in Figs. 5(e) and 5(f). The comparison of four terahertz radiation images clearly indicates that ferroelectric domains are not uniform in the depth direction and DWs are easily introduced in surface regions of the crystal.

Here, we briefly discuss the possible reason why such a ferroelectric multi-domain state appears in as-grown crystals. Single crystals of [D-55DMBP][Dia] were prepared by recrystallization at room temperature [19], which is lower than $T_c$ = 335 K. Thus, ferroelectric domain patterns are mainly determined by the kinetics of the crystal growth. In general, charged (head-to-head or tail-to-tail) DWs are electrically unstable [2], so that the presence of charged DWs is attributed to the existence of charged impurities and/or defects during the crystal growth. These impurities and defects can compensate electrical charges of DWs, resulting in ferroelectric multi-domain states. In the case of croconic acid, charged tail-to-tail DWs are only discerned in the virgin state of as-grown crystals and diminish by cyclic application of electric fields [17], which may be explained by the removal of DW pinning by external electric fields.

So far, various methods were used to observe ferroelectric domains and DWs. The most important advantage of the terahertz radiation imaging is that orientations of ferroelectric domains can be determined by phase information of terahertz waves, which is very sensitive to the direction of ferroelectric polarizations. In the present work, we successfully visualized ferroelectric domains and DWs in the inside and surface regions of as-grown crystals by using the anisotropy of the optical properties in the terahertz frequency regions. In organic molecular ferroelectrics, collective modes show up along the direction of the ferroelectric polarization [21, 34], resulting in the large optical anisotropy in the terahertz frequency region. Thus, the present method is not restricted to [D-55DMBP][Dia] but can be applied to a variety of organic ferroelectrics with optical anisotropy in the terahertz frequency region. In addition, terahertz radiation imaging is a non-destructive and contact-free method. This feature enables us to easily detect ferroelectric domain topology and DW dynamics under external electric fields (Sect. III D).

**D. Ferroelectric domain dynamics under external electric field**

In this section, we discuss ferroelectric domain dynamics under external electric fields. In this experiment, we used another as-grown crystal. The thickness of the sample was 150 μm. An optical image of the sample with two electrodes is shown in Fig. 6(a). Terahertz radiation images were taken in the area indicated by the solid white box; the actual area available for the ferroelectric domain imaging is surrounded by the dotted while lines. First, $E_{ex}$ = 10 kV/cm, which exceeds the coercive field (~2 kV/cm) [19], was applied along $2c\text{-}b$ axis in order to make the crystal a single-domain. We measured three terahertz radiation images in the front surface region with ~29 μm, inside region, and back surface region with ~29 μm, which are shown in Figs. 6(b1), 6(b2), and 6(b3), respectively. As can be seen in those figures, the crystal is occupied with a single domain with almost uniform polarization, except for the right upper region (indicated in blue) in the front surface. We show in Fig. 6 variation of the domain pattern when $E_{ex}$ (kV/cm) was changed as 10 → 0 → −10 → 0 → 10. Full domain images are compiled in the Supplemental Materials [35]. At $E_{ex}$ = 1.0 kV/cm, ferroelectric domain with a size ~100 μm (indicated in blue) is generated in the upper part of the front surface region, as shown in Fig. 6(e1). On the other hand, the single-domain state is preserved in the inside and back surface regions, which can be seen in Figs. 6(e2) and 6(e3), respectively. Further decreasing $E_{ex}$, the ferroelectric domain initially generated in the front surface region grows in size [Fig. 6(f1)] and merges into another ferroelectric domain located at the right-hand side of the crystal [Fig. 6(g1)]. When $E_{ex}$ is decreased to zero, the red-colored domain is preserved in the inside and back surface regions, which are discerned in Figs. 6(j2) and 6(j3), respectively. These results clearly indicate the appearance of the blue-colored ferroelectric domain is limited within the front surface region. By applying negative $E_{ex}$, polarization switching proceeds as follows. First, an additional ferroelectric domain indicated in blue is generated in the front surface region [Fig. 6(k1)] with uncharged 180° DWs. These DWs move along $b$-axis and diminish the original red-colored ferroelectric domain [Fig. 6(l1)]. By taking into account the anisotropy of intralayer and interlayer electrostatic interactions of molecular layers, as is discussed later in detail, these DWs are probably quasi-1D in each molecular layer parallel to the (100) plane. In the back surface region, on the other hand, the original ferroelectric domain is unchanged [Fig. 6(l3)]. Thus, the uncharged 180° DWs are generated and move only in the front region. In the inside [Fig. 6(l2)], the magnitude of the terahertz radiation is decreased to nearly zero in all area of the crystal, compared to that shown in Fig. 6(b2). This can be explained by the destructive interference of the terahertz waves radiated in the opposite ferroelectric domains in the front and back halves. In other words, uncharged 2D 180° DW probably almost parallel to the sample surface, i.e., (100) plane, is generated; the uncharged 2D 180° DW divides the crystal into two ferroelectric domains with opposite polarizations in the depth direction. The signs of $E_{THz}(0)$ are different in the back and front surface regions [Figs. 6(m1) and 6(m3)], but their magnitudes are almost identical. This indicates that this uncharged 2D 180° DW locates the middle region of the crystal, as illustrated in Fig. 7(c2). At $E_{ex}$ = −0.6 kV/cm, the front surface region is occupied with a single domain [Fig. 6(m1)]. With further increasing $|E_{ex}|$, a portion of the red-colored domains inside the crystal becomes smaller than that of the blue-colored domains [Figs. 6(o2) and 6(p2)], whereas the red-colored domain in the back surface region is almost preserved. At $E_{ex}$ = −1.2 kV/cm, uncharged 180° DW and charged head-to-head (or tail-to-tail) DW are generated in the back surface region [Fig. 6(o3) and 6(p3)]. Finally, the charged head-to-head (or tail-to-tail) DW disappears and a pair of the uncharged 180° DWs proceeds along $b$-axis, resulting in the blue-colored single domain state [Fig. 6(r3)].

When $E_{ex}$ (kV/cm) was further changed as −10 → 0 → 10, overall behavior of the ferroelectric domains [Figs. 6(r1) to 6(H1)] are almost opposite to that in the $E_{ex}$ (kV/cm) scan (10 → 0 → −10). In this process, uncharged 180° DWs firstly appear in the back surface region [Figs. 6 (s3) to 6(y3)] and propagate along $b$-axis. At 0.6 kV/cm, the red-colored single-domain state is realized in the back surface region [Fig. 6(C3)], whereas the blue-colored domain is preserved in the front surface region [Fig. 6(C1)]. Wide white-colored region is observed inside the crystal [Fig. 6C(2)], indicating the presence of uncharged 2D 180° DW almost parallel to the (100) plane. With increasing $E_{ex}$, the magnitude of the red-colored domain increases inside the crystal [Fig. 6(D2)] and uncharged 180° DW shows up in the front surface region [Fig. 6(D1)]. Finally, uncharged 180° DW moves along *b*-axis, resulting in the red-colored single domain in the whole volume [Figs. 6(H1) to 6(H3)].

To clearly see the domain switching in response to $E_{ex}$, we show in Fig. 7(a) the $E_{ex}$ dependence of the total amplitude of the terahertz radiation, $E_{THz}^{area}$, obtained by the integration of $E_{THz}(0)$ over the entire crystal area in the $E_{ex}$ (kV/cm) scan (10 → 0 → −10 → 0 → 10); the red, black, blue circles indicate $E_{THz}^{area}$ in front surface, inside, and back surface regions. $E_{THz}^{area}$ from the inside of the crystal shows a hysteresis loop, reflecting the bulk ferroelectric nature. Furthermore, two kinks can be discerned at around ±0.5 kV/cm. Compared to $E_{THz}^{area}$ in the inside region, the hysteresis loops of $E_{THz}^{area}$ in the front and back surface regions are shifted by ~0.6 kV/cm to opposite direction. Namely, the coercive field of the front surface region is different from that of the back surface region. The hysteresis loop of $E_{THz}^{area}$ in the inside images appears to be reproduced by the average of those of the front and back halves. This implies that DW motion in the each half rules the domain switching behavior.

In order to discuss the stability of DW under $E_{ex}$, we estimate the position of a DW along the depth direction, i.e., *Z*-direction, using terahertz radiation images in front, inside, and back surface regions (Fig. 6); *X*-, *Y*-, and *Z*-directions relative to crystal orientations are illustrated in the left side of Fig. 7(b). The crystal becomes single-domain states under $E_{ex} = \pm 10$ kV/cm [Figs. 6(b), 6(r), and 6(H)], while the crystal is divided into two domains by an uncharged 2D 180° DW under $E_{ex} = -0.6$ kV/cm [Fig. 6(m)] and $E_{ex} = 0.6$ kV/cm [Fig. 6(C)]. This allows us to assume only one domain boundary in the depth direction; DW depth is 0 μm and 150 μm under $E_{ex} = +10$ kV/cm and $E_{ex} = -10$ kV/cm, respectively. Under this assumption, $E_{THz}(0)$ in the terahertz radiation images in the inside [Figs. 6(b2)-6(r2)] normalized by that in $E_{ex} = +10$ kV/cm [Fig. 6(r2)] correspond to the DW depth $z$, $z$ is estimated by the relation $E_{THz}(0) = -\int_0^z G(z')dz' + \int_z^d G(z')dz'$. The DW was projected on the Y-Z plane by averaging $z$ in the X-direction (∥ 2c-b) for simplicity. Figure 7(b) shows the $E_{ex}$ dependence of the DW depth when $E_{ex}$ was changed as +10 → 0 → −10; $z$ = 0 μm and $z$ = 150 μm correspond to positions of the front and back surfaces, respectively. At $E_{ex}$ = 1.6 kV, a DW shows up in the upper region of the front surface region, as discerned by the red line in Fig. 7(b). The DW is not flat but tilted in the Y-Z plane. Figure 7(c1) shows the schematic illustration of prospected position of the DW (indicated by the gray area); arrows indicate the direction of the ferroelectric polarization. At $E_{ex}$ = –0.2 kV/cm, the position of DW locates at $z$ ~ 70 μm and the DW became flat, as indicated by the green line in Fig. 7(b). This corresponds to the completion of domain switching in the front half as schematically shown in Fig. 7(c2). In the back half, a tilted DW appears again at $E_{ex}$ = –1.4 kV/cm [the blue line in Fig. 7(b)], the schematic illustration of which is shown in Fig. 7(c3). With increasing |$E_{ex}$|, the tilted DW in the upper region finally diminishes at $E_{ex}$ = −10 kV/cm, which may be related to the Barkhausen pulses frequently observed in ferroelectrics [1, 36].

By taking into account the anisotropy of intralayer and interlayer electrostatic interactions of the molecular layers, we briefly discuss the reason why polarization switching proceeds with successive propagations of uncharged 180° DW along the

*b*-axis (Fig. 6) and uncharged 2D 180° DW almost parallel to the (100) plane is finally formed [Fig. 7(c)]. [D-55DMBP][Dia] is formed by hydrogen-bonded chain layers in the (100) plane and their stacking along the direction perpendicular to the (100) plane, which are schematically shown in Fig. 7(d). The precise structural data of [D-55DMBP][Dia] is not available. Thus, we used X-ray structural data of the related analog [H-55DMBP][Hia] at 50 K in the ferroelectric phase [19] to deduce the anisotropy in intralayer and interlayer electrostatic interactions. Protons in the ferroelectric phase occupy four crystallographically nonequivalent sites within hydrogen-bonded chain plane, which are labeled by H(1), H(2), H(3), and H(4) in Fig. 1. The H(1)-H(2) and H(3)-H(4) distances were estimated to be ~4.148 Å and ~5.989 Å, respectively, which are much shorter than the distance (~8.725 Å) between the neighboring layers. Thus, interlayer electrostatic interaction along the stacking direction, i.e., perpendicular to the (100) plane, should be weak, compared to intralayer electrostatic interaction. This causes the formation of uncharged intralayer 1D 180° DW within the hydrogen-bonded chain layers. By the application of $E_{ex}$ along the 2*c*-*b* axis, transfer of protons with the hydrogen-bonded chain direction occurs so that intrachain domain boundary [the dotted line in Fig. 7(e1)] separated by a head-to-head (or tail-to-tail) polarization displaces along the 2*c*-*b* axis [Figs. 7(e1) and 7(e2)]. Subsequently, the uncharged intralayer 1D 180° DW [the dotted line in Fig. 7(e2)] moves along the *b*-axis [Figs. 7(e3) and 7(e4)], which is consistent with the observed DW dynamics (Fig. 6). This process ideally causes uncharged interlayer 180° DWs under nearly zero electric fields [the dotted line in Fig. 7(d2) and 7(c2)], which agrees well with the observation of uncharged 2D 180° DW almost parallel to the (100) plane. However, actual DW is stabilized by subtle competition between intralayer and interlayer electrostatic interactions so that it is governed by the presence of pinning centers such as defects induced during the crystal growth. This would be the main origin of the observed tilted DW [Figs. 7(c1) and 7(c3)].

## IV. Conclusion

We successfully visualized the ferroelectric domain topology of a room-temperature organic supramolecular ferroelectric [D-55DMBP][Dia] by terahertz radiation imaging. Using the anisotropy in the effective depth of the region contributing to the terahertz waves, we developed a new method to detect the ferroelectric domains and domain walls (DWs) in the inside and surface regions of the as-grown crystal. Application of the external electric field along the $2c$-$b$ axis caused the generation of an uncharged 2D 180° DW almost parallel to the (100) plane. The domain switching process is explained by the propagation of the uncharged intralayer 1D 180° DW along the *b*-axis. This new method presented here can be applied to quasi-three-dimensional visualization of ferroelectric domains and DWs in various organic ferroelectrics with optical anisotropy in the terahertz frequency region.

## Acknowledgements

This work was partly supported by a Grant-in-Aid by MEXT (No. 25247049, 25600072, and No. 25-3372). M. S. was supported by Japan Society for the Promotion of Science (JSPS) through Program for Leading Graduate Schools (MERIT) and JSPS Research Fellowships for Young Scientists.

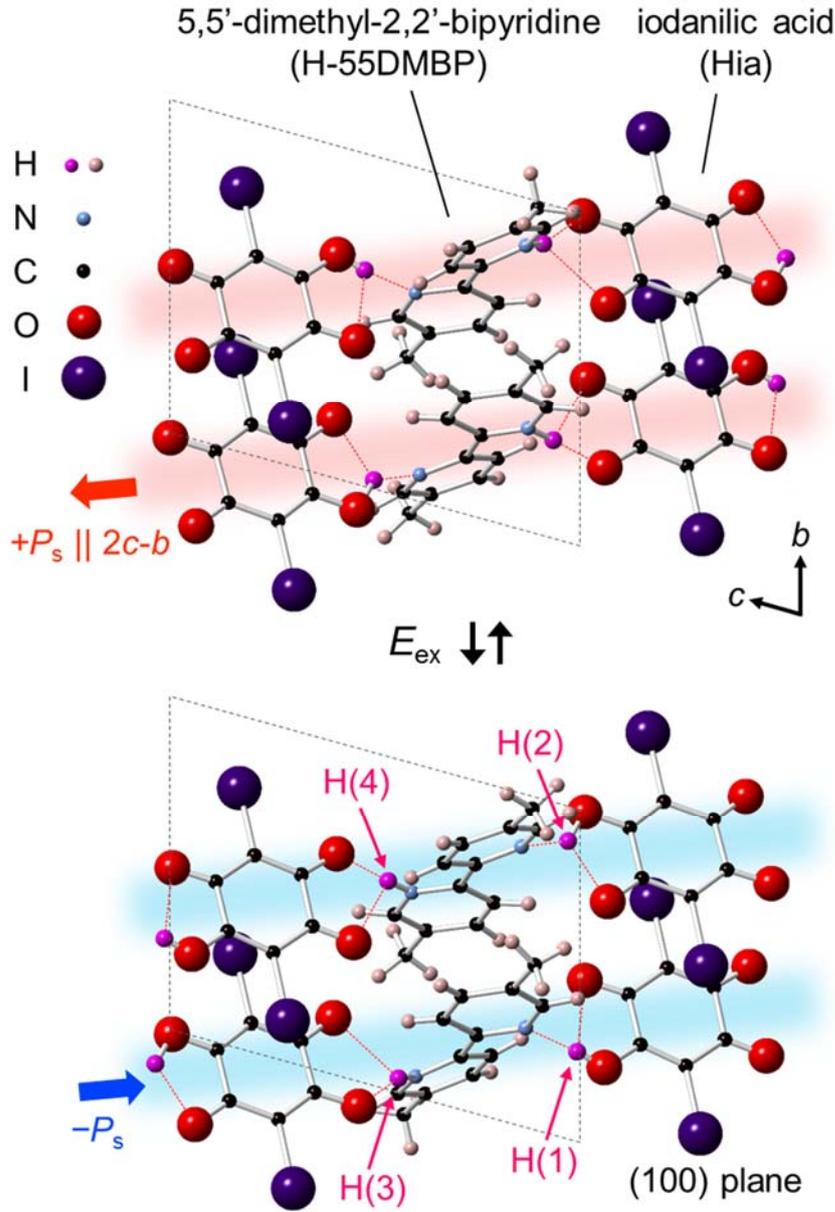

FIG. 1. (Color online) Schematic illustration of the crystal structure of [H-55DMBP][Hia] in the ferroelectric phase (50 K) [19]. Dashed parallelogram indicates the unit cell. Spontaneous polarization $P_s$ shows up along the $2c$-$b$ axis and polarization switching is induced by the application of external electric field $E_{ex}$.

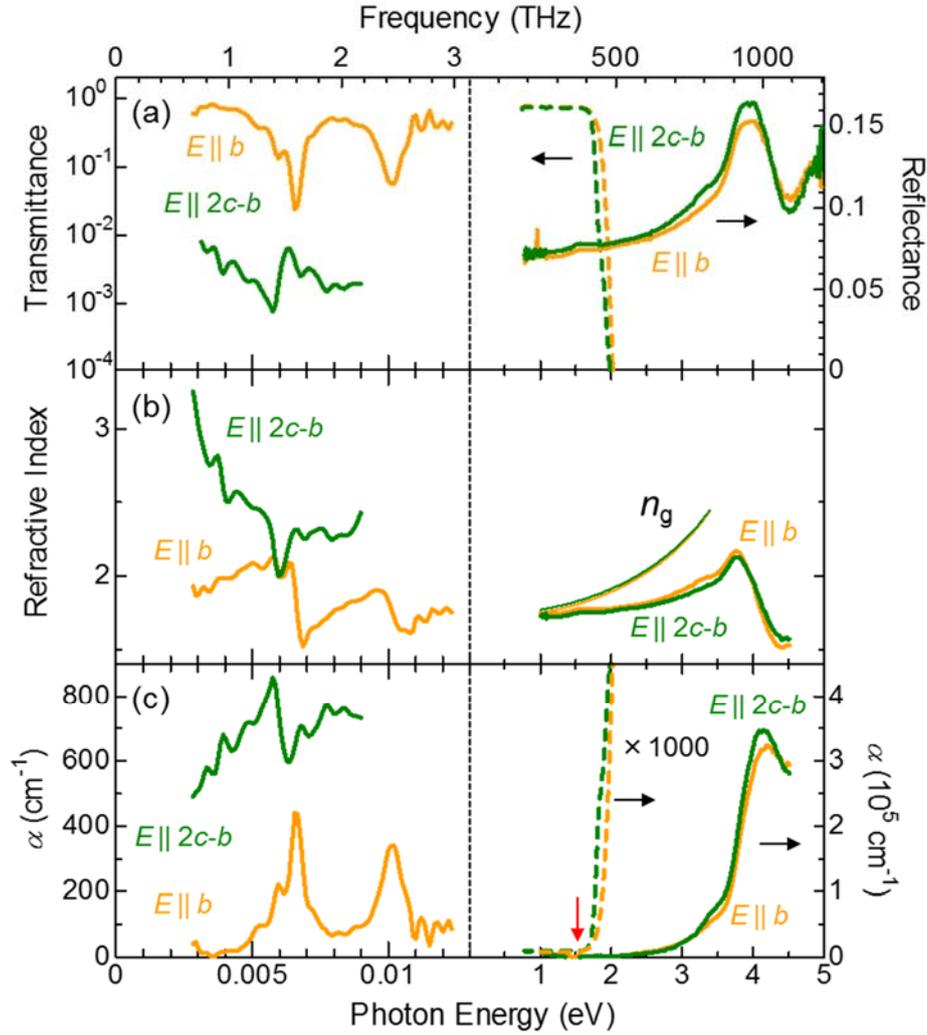

FIG. 2. (Color online) Polarized optical spectra of [D-55DMBP][Dia] for $E||2c-b$ (orange lines) and $E||b$ (green lines) in terahertz and visible frequency regions, measured at room temperature. (a) Transmittance and reflectance, (b) refractive index, and (c) Absorption coefficient $\alpha$ spectra. We also show in (b) the group refractive index $n_g$ spectrum derived from the Sellmeier relationship. Dotted and solid lines in (c) were calculated from the transmittance spectra and the Kramers-Kronig transformation of the reflectance, respectively. The red downward arrow in (c) indicates the photon energy of the femtosecond laser pulse used in terahertz radiation experiments.

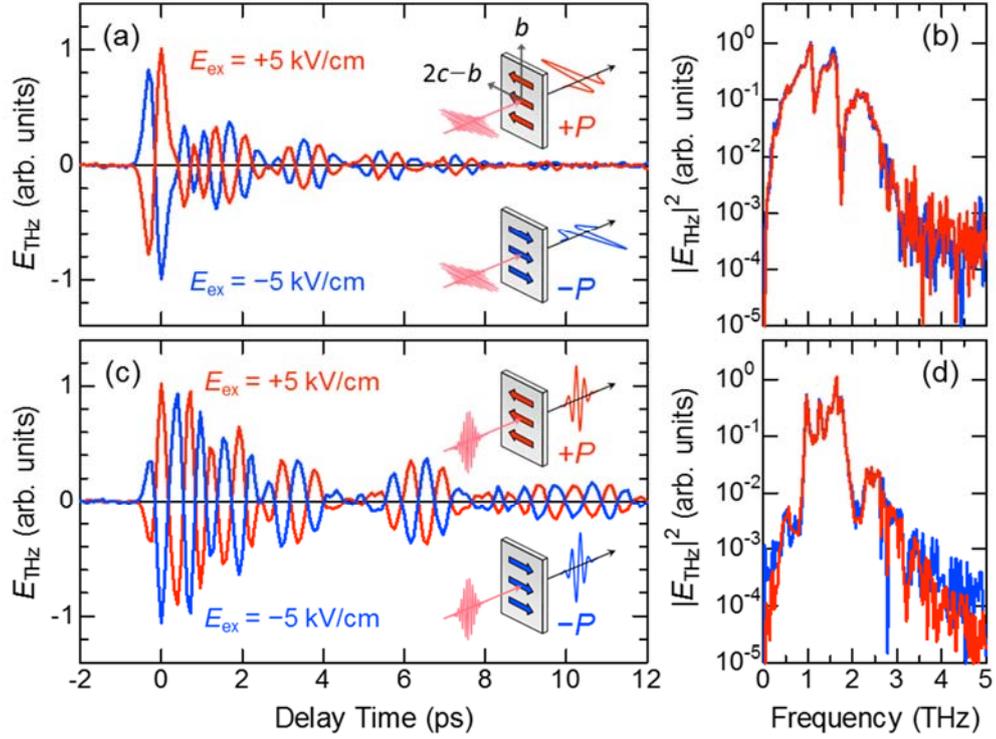

FIG. 3. (Color online) (a) Electronic-field waveforms emitted from [D-55DMBP][Dia] in $E^{\omega}||2c-b$ and $E_{\text{THz}}||2c-b$ configurations, measured at room temperature. We applied the external electric field $E_{\text{ex}}$ of ±5 kV/cm, which exceeds the coercive field (~2 kV/cm). (b) Fourier-transformed power spectra. (c) Electric-field waveforms and (d) power spectrum in $E^{\omega}||b$ and $E_{\text{THz}}||b$ configurations. Insets in (a) and (c) show schematic illustration of the optical configurations with respect to crystal orientation. Electric polarization $P$ is represented by the arrow.

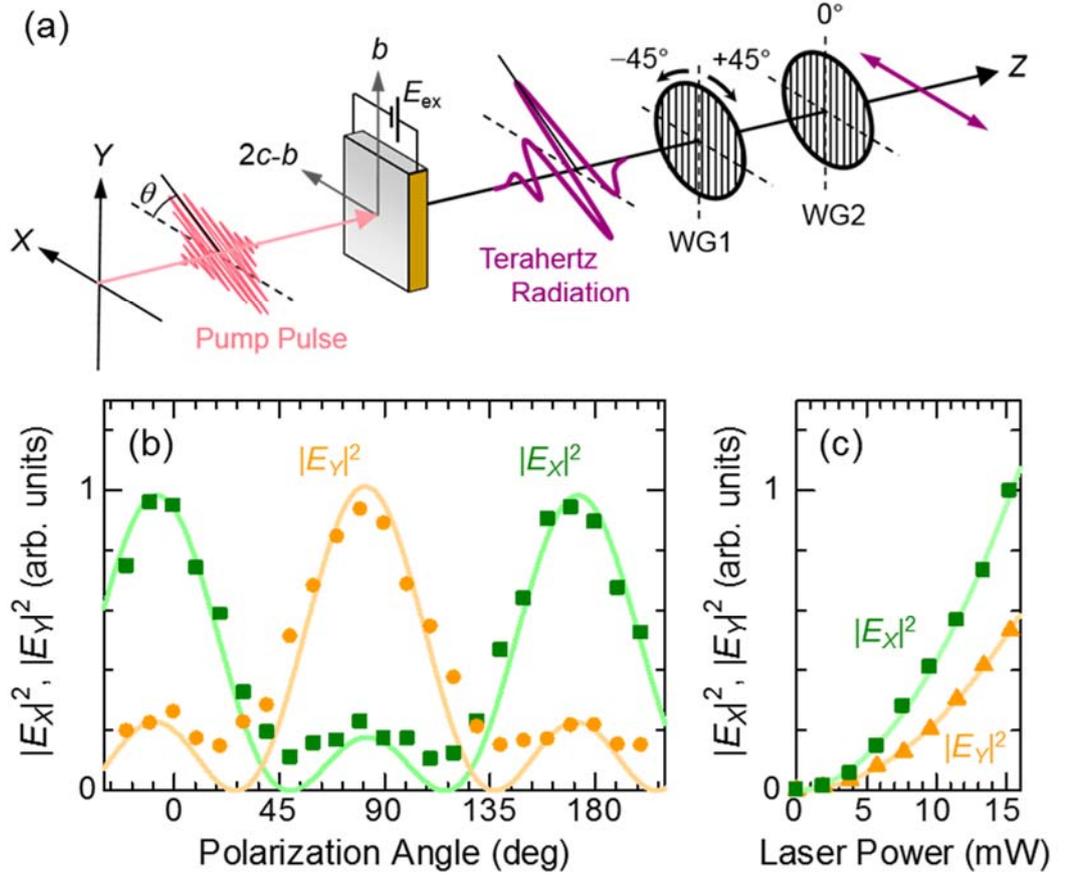

FIG. 4. (Color online) (a) Schematic illustration of the experimental setup. We used two wire grids (WGs); the angle of WG1 was set to 45° or -45° with respect to the $Y$-direction. The external electric field $E_{ex}$=+5 kV/cm was applied. (b) Laser polarization dependence of the intensity of the terahertz waves $|E_{THz}|^2$ in the frequency range of 0.8-1.2 THz. Solid lines are least-square fittings by Eq. (4). (c) Laser power dependence of $|E_{THz}|^2$ with $\theta = 0°$. Solid lines are quadratic fitting curves.

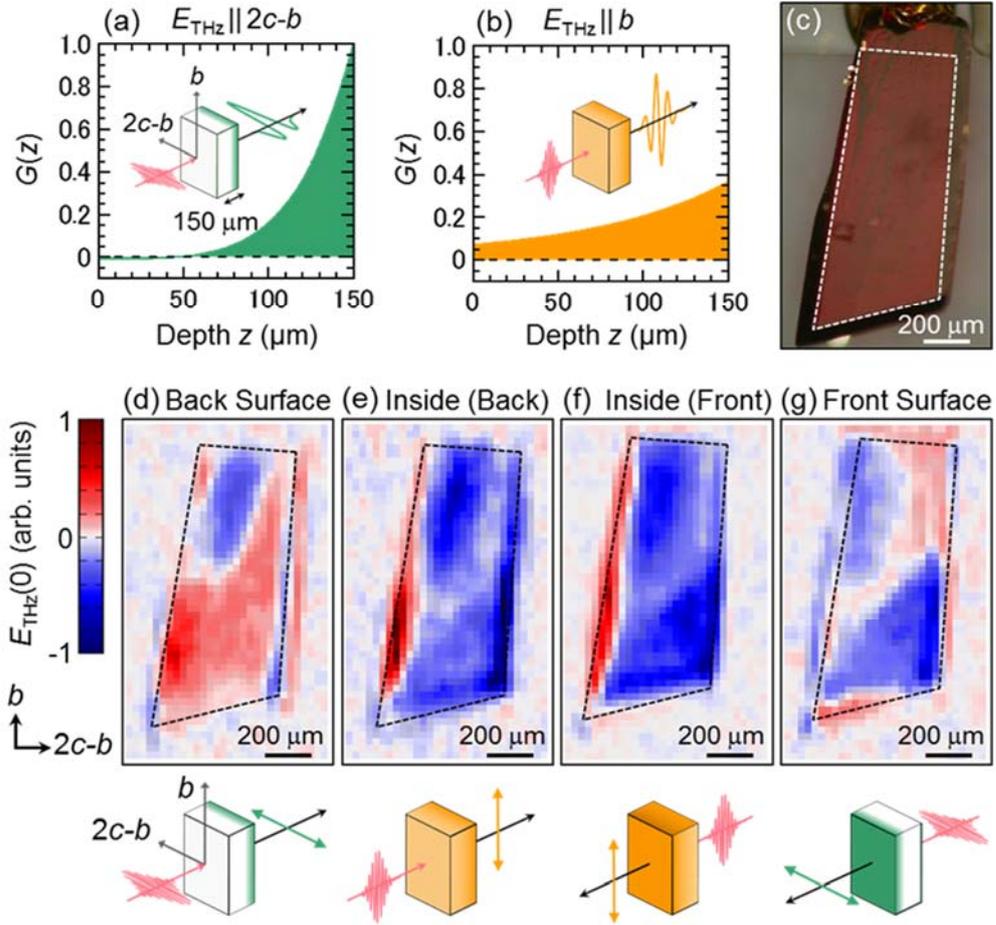

FIG. 5. (Color online) (a) Depth profile of the terahertz radiation $G(z)$ in $E^\omega \| 2c - b$ and $E_{\text{THz}} \| 2c - b$ configurations with a sample thickness $d$ of 150 μm. (b) $G(z)$ in $E^\omega \| b$ and $E_{\text{THz}} \| b$ configurations. $G(z)$ is normalized to give the same area in (a) and (b). Insets in (a) and (b) show the optical configurations. (c) Optical microscope image. (d-g) Terahertz radiation images in the optical configurations depicted below. Dashed square in (c-g) indicate the valid area of the terahertz radiation images.

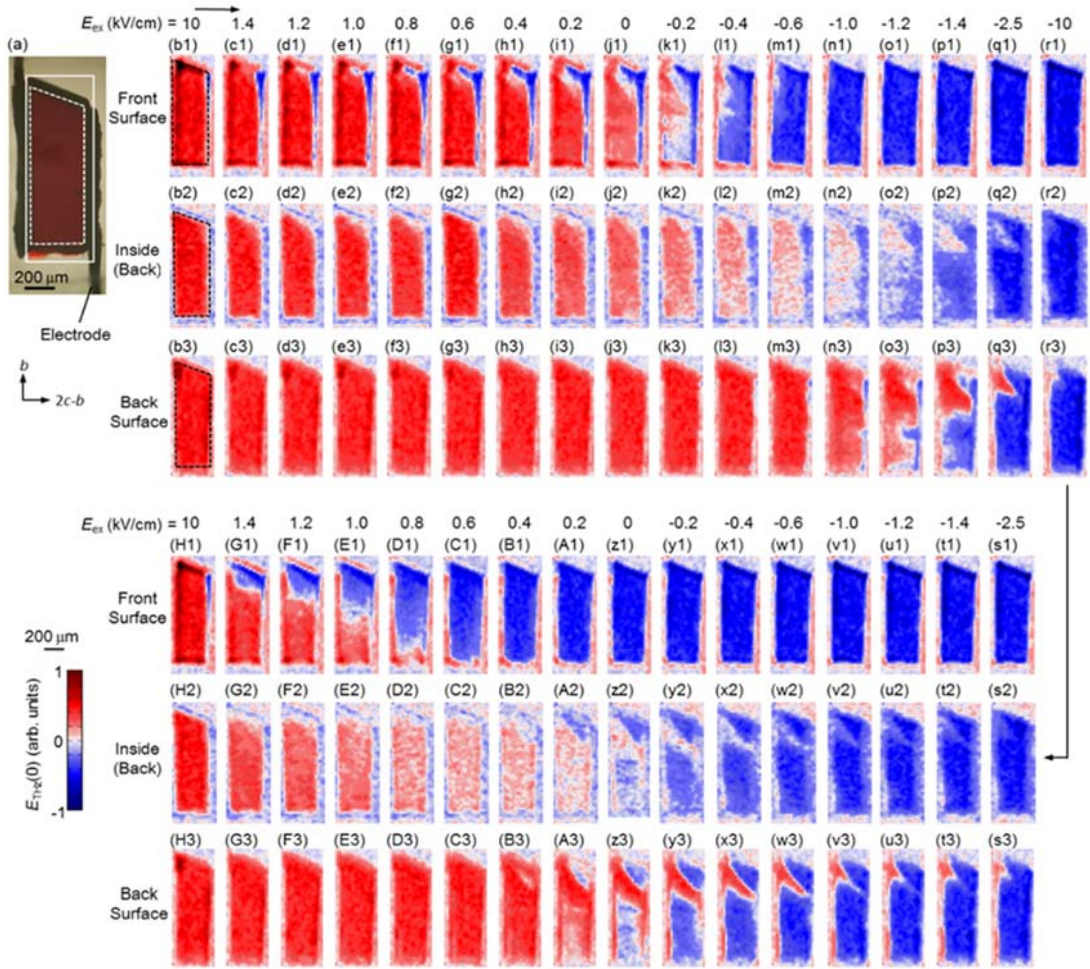

FIG. 6. (Color online) (a) Optical microscope image. (b-H) Terahertz radiation images in (1) front surface, (2) inside, and (3) back surface regions under electric fields. Full data are compiled in the Supplemental Movie [35].

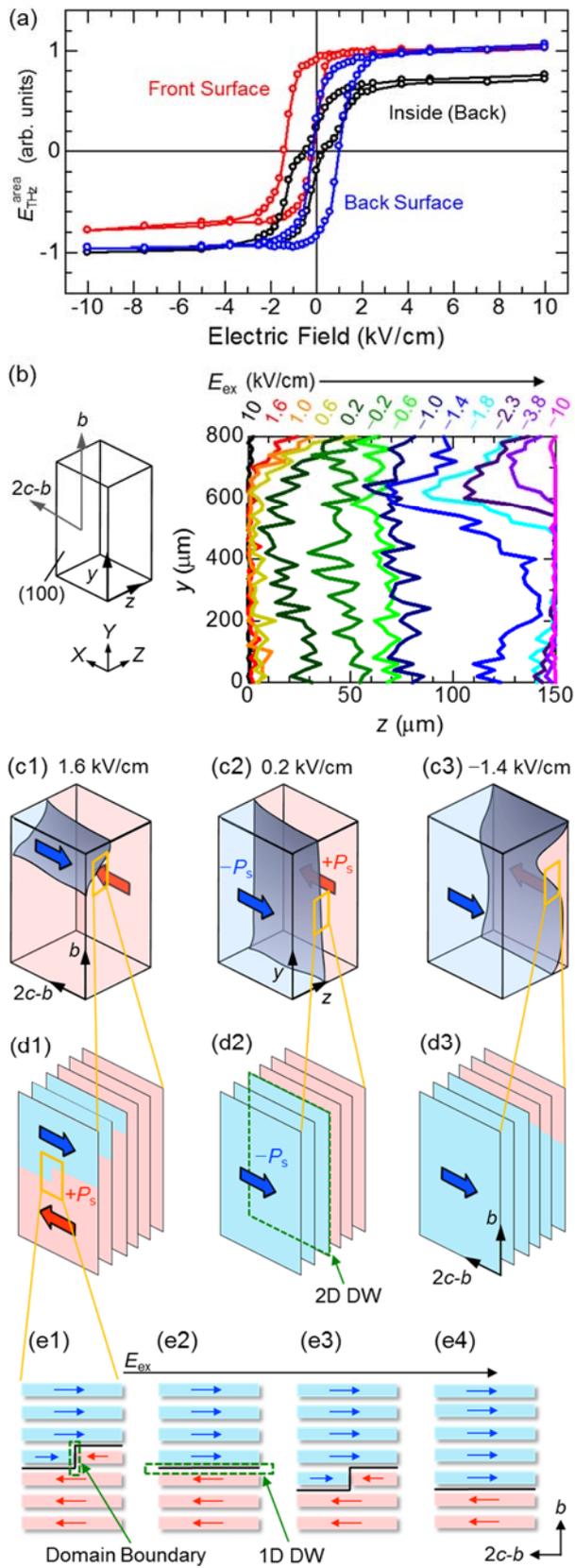

FIG. 7. (Color online) (a) Electric field dependence of the total amplitude of the images in front surface, inside, and back surface regions in Fig. 6. (b) Estimated DW depth on the *Y-Z* plane in 10 kV/cm → −10 kV/cm. *X*-, *Y*-, and *Z*-axes relative to crystal orientations are schematically shown. Schematic illustrations focused on (c) DWs and (d) molecular stacking. 2D DW is surrounded by the dotted line in (d2). (e) Microscopic interpretation of the ferroelectric polarization switching process under electric fields. Domain boundary and 1D DWs are surrounded by the dotted lines.